\documentclass[]{piparticle-final}

\usepackage{graphicx}
\usepackage{amsmath}
\usepackage{cite} 
\usepackage{epstopdf}

\usepackage{booktabs}
\usepackage{listings,caption,subcaption,float}

\begin{document}

\volume{7}               
\articlenumber{070018}   
\journalyear{2015}       
\editor{R. Dickman}   
\reviewers{M. Hutter, Australian National University, Canberra, Australia.}  
\received{2 November 2015}     
\accepted{27 November 2015}   
\runningauthor{D. J. R. Sevilla}  
\doi{070018}         

\title{Bayesian regression of piecewise homogeneous Poisson processes}

\author{Diego J. R. Sevilla\cite{inst1}\thanks{E-mail: dsevilla@fceia.unr.edu.ar}  
}

\pipabstract{
In this paper, a Bayesian method for piecewise regression is adapted to handle counting processes data distributed as Poisson.
A numerical code in \emph{Mathematica} is developed and tested analyzing simulated data.
The resulting method is valuable for detecting breaking points in the count rate of time series for Poisson processes.
}
                                                                                    
\maketitle

\blfootnote{
\begin{theaffiliation}{99}
   \institution{inst1}
Departamento de F\'{\i}sica y Qu\'{\i}mica, Escuela de Formaci\'on B\'asica.
Facultad de Ciencias Exactas, Ingenier\'{\i}a y Agrimensura.
Universidad Nacional de Rosario, Av. Pellegrini 250, S2000BTP Rosario, Argentina.
\end{theaffiliation}
}

\section{Introduction}\label{sec:intro}

Bayesian statistics have revolutionized data analysis \cite{gre05}.
Techniques like the Generalized Lomb-Scargle Periodogram \cite{bre88}
allow us to obtain oscillation frequencies of time series with unprecedented accuracy.
The Gregory and Loredo method \cite{gre92} goes
further allowing us to find and characterize periodic signals of any period
and shape.

To detect non-periodical variations, the \emph{Exact Bayesian
Regression of Piecewise Constant Functions} by Marcus Hutter
(hereafter Hutter's method) \cite{hut07} 
is valuable.  It permits to estimate the most probable partition of a
data set in segments of constant signals, determining the number of
segments and their borders, and in-segments means and variances.
Hutter's method works with two continuous distributions: Normal,
and Cauchy-Lorentz.  The latter ---the canonical example of a
pathological distribution with undefined moments---, is also suitable to
analyze data with other symmetric probability distributions,
especially with heavy tails.

In the case of counting processes, especially for low rates, when data
consist in non-negative small integers, methods specially designed to
discrete probability distributions are necessary.  Some regression
methods, specially for non-homogeneous Poisson processes \cite{law87},
were developed.

In this paper, Hutter's method is adapted for analyzing data
distributed as Poisson.  The results are summarized in a code in
\emph{Mathematica} \cite{mat12}.  It can be used to analyze data of
several physical processes which follow the Poisson distribution
(e.g., detection of photons in X-ray Astronomy, particles in nuclear
disintegration, etc.), if sudden changes in detection rates are
suspected.

\section{Method}

Hutter's method is summarized in Table 1 of Ref. \cite{hut07} in
a pseudo C code which is divided in two blocks.  The first one
calculates moments $A^k_{ij}$ with $k=0,1,2$ of the PDF of the
statistical models for segments of data
$D_{ij}:=\{n_{i+1},\ldots,n_j\}$.  The second one performs the
regression from moments $A^k_{ij}$.  The code developed in this work
is divided in three blocks.

As the members of the Poisson distributions family are identified by
one parameter -the mean rate $r$ of the Poisson process-, the PDF of
the models for a segment $D_{ij}$ is \cite{gre05}
\begin{equation}
P(r|D_{ij},I)=\frac{P(r|I)\,P(D_{ij}|r)}{P(D_{ij})},
\end{equation}
where $P(r|I)$ is the prior of parameter $r$, $P(D_{ij}|r)$ is the
likelihood of segment $D_{ij}$ for a given $r$, $P(D_ {ij})$ is the
global likelihood of the family, and $I$ represents a prior
information.

Usually, the prior information consists of global quantities calculated from
$D:=D_{0N}$, \emph{i.e.}, from all the data set. For Poisson
processes, only one quantity is necessary: the mean rate
$\hat{r}$. Considering the conjugate prior of the Poisson distribution
\cite{gel14}, the prior results
\begin{equation}
P(r|\hat{r})=\frac{r^{\hat{r}-1}\,e^{-r}}{\Gamma(\hat{r})}.
\end{equation}

For a Poisson process with rate $r$, the likelihood of a segment $D_{ij}$ is
\begin{equation}
P(D_{ij}|r)=\prod_{t=i+1}^j \frac{r^{n_t}\,e^{-r}}{n_t!}.
\end{equation}

So, the moments of the posterior
can be expressed in an analytical form
\begin{equation}
A^k_{ij}=\frac{\Gamma(k+\hat{r}+\sum_{t=i+1}^j n_t)\,\prod_{t=i+1}^j \frac{1}{n_t!}}{\Gamma(\hat{r})\,(j-i+1)^{k+\hat{r}+\sum_{t=i+1}^j n_t}}.
\end{equation}

Code block \ref{prog1} calculates $A^k_{ij}$.  It needs as input the
time series to be analyzed (list {\ttfamily{data}}).  The output are
functions {\ttfamily{A0[i,j]}}, {\ttfamily{A1[i,j]}} and
{\ttfamily{A2[i,j]}} and integer {\ttfamily{n}}, which is the length
of {\ttfamily{data}}.

\begin{lstlisting}[frame=single,basicstyle=\ttfamily\scriptsize,language=Mathematica,caption=\emph{Mathematica} code to calculate $A^k_{ij}$.,label=prog1,morekeywords={IntegerPartitions}]
n=Length[data];
r=Mean[data];
Do[
  Do[
    d=j-i; m=r+Sum[data[[t]],{t,i+1,j}];
    A0[i,j]=(m-1)!/(Gamma[r]*(d+1)^m)*
      Product[1/data[[t]]!,{t,i+1,j}];
    A1[i,j]=m*A0[i,j]/(d+1);
    A2[i,j]=(m+1)*A1[i,j]/(d+1);
  ,{j,i+1,n}];
,{i,0,n}];
\end{lstlisting}

As the second block of Hutter's code only needs the moments $A^k_{ij}$
as inputs, it could work properly with no changes.  It computes the
evidence, the probability for $k$ segments and its MAP estimation
$\hat{k}$, the probability of boundaries locations and the MAP
locations of the $\hat{k}$ boundaries, the first and second in-segment
moments, and an interesting regression curve that smooths the final
result.

Nevertheless, for our specific problem, once the segments boundaries are obtained,
we can estimate their means and variances
straightforwardly, so we only use a part of Hutter's second block,
which is shown in code block \ref{prog2}.  The logical of the
algorithm is explained in Ref. \cite{hut07}.  Code block
\ref{prog2} needs as inputs {\ttfamily{A0[i,j]}},
{\ttfamily{A1[i,j]}}, {\ttfamily{A2[i,j]}} and {\ttfamily{n}}, all
calculated in code block \ref{prog1}, and integer {\ttfamily{kmax}},
which is the maximum number of segments to be considered.  The outputs
are the evidence ({\ttfamily{e}}), the probability for $k$ segments
({\ttfamily{c[k]}}), its MAP ({\ttfamily{khat}}), the probability of
boundaries locations ({\ttfamily{B[i]}}), and their MAP
({\ttfamily{that[p]}}).

\begin{lstlisting}[frame=single,basicstyle=\ttfamily\scriptsize,language=Mathematica,caption=\emph{Mathematica} code to calculate breaking points.,label=prog2,morekeywords={IntegerPartitions}]
Do[
  L[0,i]=KroneckerDelta[i,0];
  R[0,i]=KroneckerDelta[i,n];
,{i,0,n}];
Do[
  Do[
    L[k+1,i]=Sum[L[k,h]*A0[h,i],{h,k,i-1}];
    R[k+1,i]=Sum[R[k,h]*A0[i,h],{h,i+1,n-k}];
  ,{i,0,n}]
,{k,0,kmax-1}];
e=1/kmax*Sum[L[k,n]/Binomial[n-1,k-1],{k,1,kmax}];
Do[
  c[k]=L[k,n]/(Binomial[n-1,k-1]*kmax*e)
,{k,1,kmax}];
khat=1;
Do[If[c[khat]<c[k],khat=k];,{k,0,kmax}];
Do[B[i]=
  Sum[L[p,i]*R[khat-p,i]/L[khat,n],{p,0,khat}];
,{i,0,n}];
Do[
  that[p]=0;
  dummy=L[p,0]*R[khat-p,0];
  Do[
    If[dummy<L[p,h]*R[khat-p,h],
      that[p]=h;
      dummy=L[p,h]*R[khat-p,h];
    ];
  ,{h,1,n}];
,{p,0,khat}];
\end{lstlisting}

Finally, we calculate the in-segments means and estimate their
statistical errors.  If one segment of $m$ elements has $n$ counts,
its mean rate is $n/m$, and its variance is approximately $n/m^2$.

To determine the accuracy of the fit, it is also useful to estimate
the uncertainties of the boundaries locations.  A reasonable
estimation of the uncertainties can be obtained from the second
moments of the probability distributions of boundaries locations in
the neighborhoods of the breaking points.  If the probabilities of the
boundaries locations are given by function $B(k)$, the uncertainties
approximately result
\begin{equation}
\epsilon_{\hat{k}}=
\sqrt{\frac{
\sum_{k=\hat{k}-a+1}^{\hat{k}+a} B(k)\,(k-\hat{k})^2}{\sum_{k=\hat{k}-a+1}^{\hat{k}-a} B(k)}},
\end{equation}
where $a$ is large enough to consider all the
width of the peak
corresponding to boundary $\hat{k}$, but small enough to
avoid including peaks of other boundaries.  In practice, if $\hat{\j}$,
$\hat{k}$ and $\hat{l}$ are the positions of consecutive boundaries, a
good value for $a$ is the minimum of $(\hat{k}-\hat{\j})/2$ and
$(\hat{l}-\hat{k})/2$.

Code block \ref{prog3} uses these approaches to calculate the constant
piecewise regression of data, and to estimate its statistical errors.  The
inputs are {\ttfamily{c[k]}}, {\ttfamily{khat}}, {\ttfamily{B[i]}} and
{\ttfamily{that}}.  The outputs are list {\ttfamily{reg}}, which
represents the best fit, and lists {\ttfamily{re1}} and
{\ttfamily{re2}}, which represent the minimum and maximum estimations
considering the statistical errors of {\ttfamily{reg}}.

\begin{lstlisting}[frame=single,basicstyle=\ttfamily\scriptsize,language=Mathematica,caption=\emph{Mathematica} code to calculate regression.,label=prog3,morekeywords={IntegerPartitions}]
TBP=DeleteDuplicates[Table[that[k],{k,0,khat}]];
NBP=Length[TBP]-1;
Do[BP[k]=TBP[[k+1]];,{k,0,NBP}];
ethat[0]=0;ethat[NBP]=0;
Do[
  l1=IntegerPart[(BP[k-1]+BP[k])/2]+1;
  l2=IntegerPart[(BP[k+1]+BP[k])/2];
  ethat[k]=
    Round[Sqrt[Sum[(j-BP[k])^2*B[j],{j,l1,l2}]/
    Sum[B[j],{j,l1,l2}]]];
,{k,1,NBP-1}];
Do[
  mm[k]=BP[k]-BP[k-1];
  nn[k]=Sum[data[[i]],{i,BP[k-1]+1,BP[k]}];
,{k,1,NBP}];
reg=Flatten[Table[Table[nn[k]/mm[k]
  ,{BP[k]-BP[k-1]}],{k,1,NBP}]];
BP1[0]=BP2[0]=BP[0];BP1[NBP]=BP2[NBP]=BP[NBP];
Do[
  s=If[reg[[BP[k]+1]]>reg[[BP[k]]],1,-1];
  BP1[k]=BP[k]+s*ethat[k];
  BP2[k]=BP[k]-s*ethat[k];
,{k,1,NBP-1}];
re1=Flatten[Table[Table[(nn[k]-Sqrt[nn[k]])/mm[k]
  ,{BP1[k]-BP1[k-1]}],{k,1,NBP}]];
re2=Flatten[Table[Table[(nn[k]+Sqrt[nn[k]])/mm[k]
  ,{BP2[k]-BP2[k-1]}],{k,1,NBP}]];
\end{lstlisting}

\section{Applications and Discussion}

\begin{figure}[]
	\captionsetup[subfigure]{aboveskip=0pt,belowskip=-8pt}
	\centering
	\begin{subfigure}[b]{0.8\columnwidth}
		\hspace{0.01cm} \includegraphics[height=0.7\columnwidth]{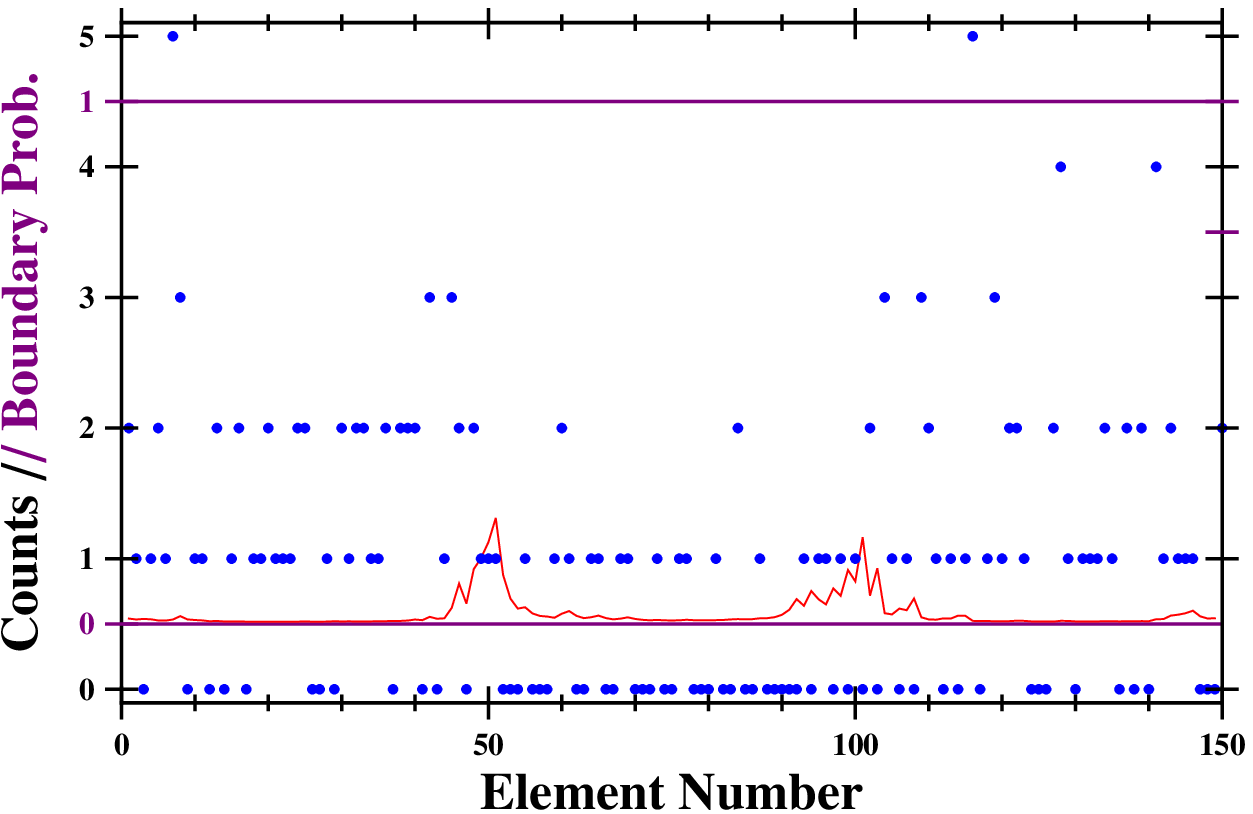}
	\end{subfigure}
	\vspace{0.2cm}
	
	\begin{subfigure}[b]{0.8\columnwidth}
		\includegraphics[height=0.7\columnwidth]{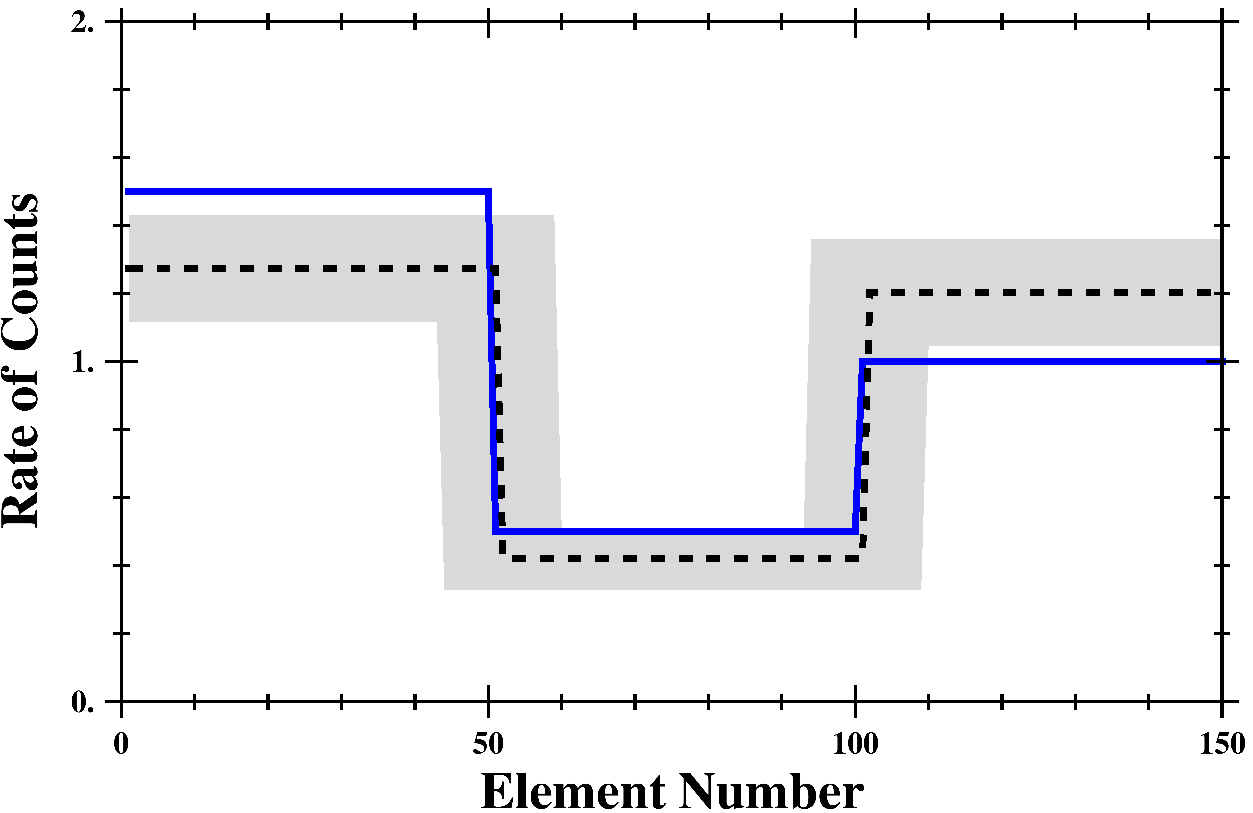}
	\end{subfigure}
	\caption{
		Top: Simulated data (blue dots) and boundaries location probability (red line).
		Bottom: Regression curve and its error estimation (black dashed line and gray zone), and the rate curve used in simulation (blue line).
	}
	\label{fig:simu1}
\end{figure}

Figure \ref{fig:simu1} (top) shows, in blue dots, data simulated using
\emph{Mathematica}.  Data consist of 150 Poisson distributed elements, the
first 50 with rate $1.5$, the second 50 with rate $0.5$, and the last
50 with rate $1.0$.  Applying the first 2 blocks of code on data, we
can see that the probability of having 2 breaking points is very high.
Figure \ref{fig:simu1} (top) also shows, in red line, the probability
for the boundaries locations.  Applying code block 3, we obtain the
regression [Fig. \ref{fig:simu1} (bottom), black dashed curve] and its error
estimation [Fig. \ref{fig:simu1} (bottom), gray zone]. The continuous blue
line in Fig. \ref{fig:simu1} (bottom) indicates the rates used in
simulation.

\begin{figure}[]
	\captionsetup[subfigure]{aboveskip=0pt,belowskip=-8pt}
	\centering
	\begin{subfigure}[b]{0.8\columnwidth}
		\hspace{0.01cm} \includegraphics[height=0.7\columnwidth]{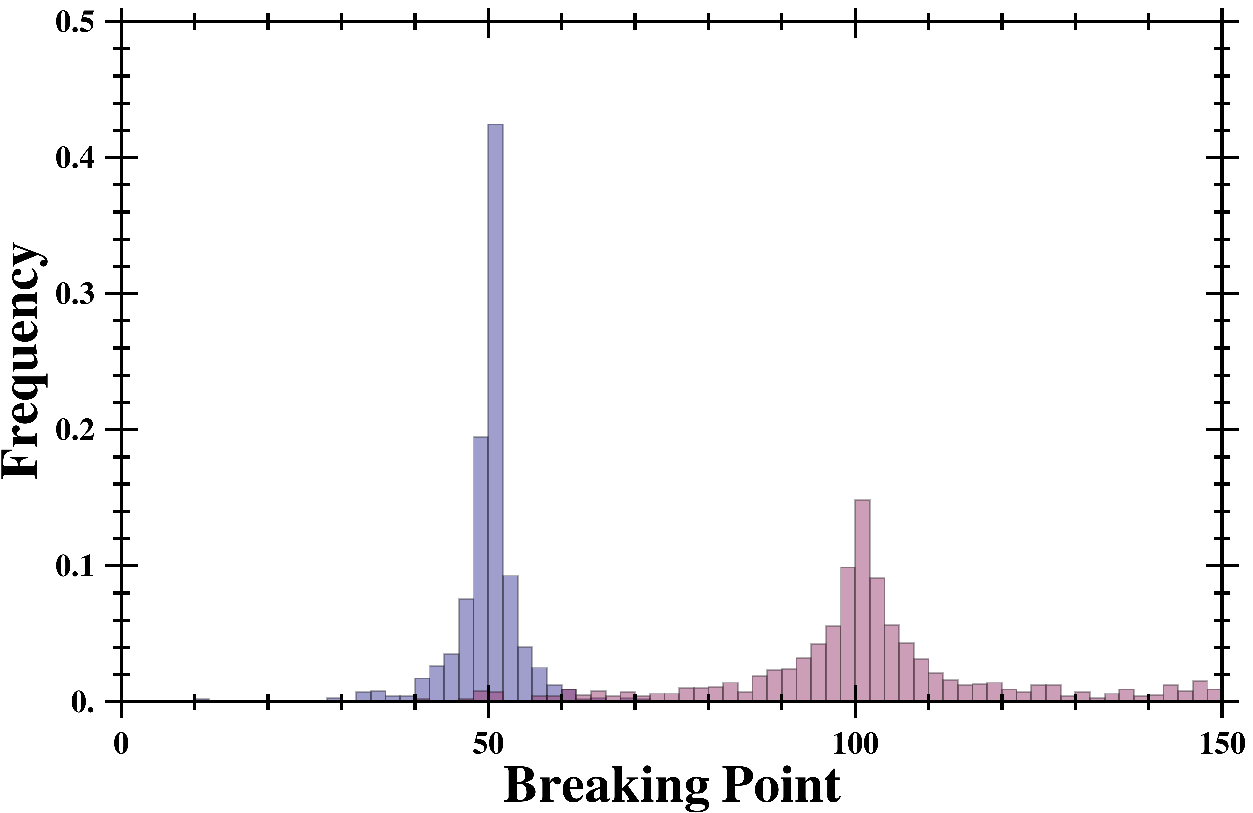}
	\end{subfigure}
	\vspace{0.2cm}
	
	\begin{subfigure}[b]{0.8\columnwidth}
		\includegraphics[height=0.7\columnwidth]{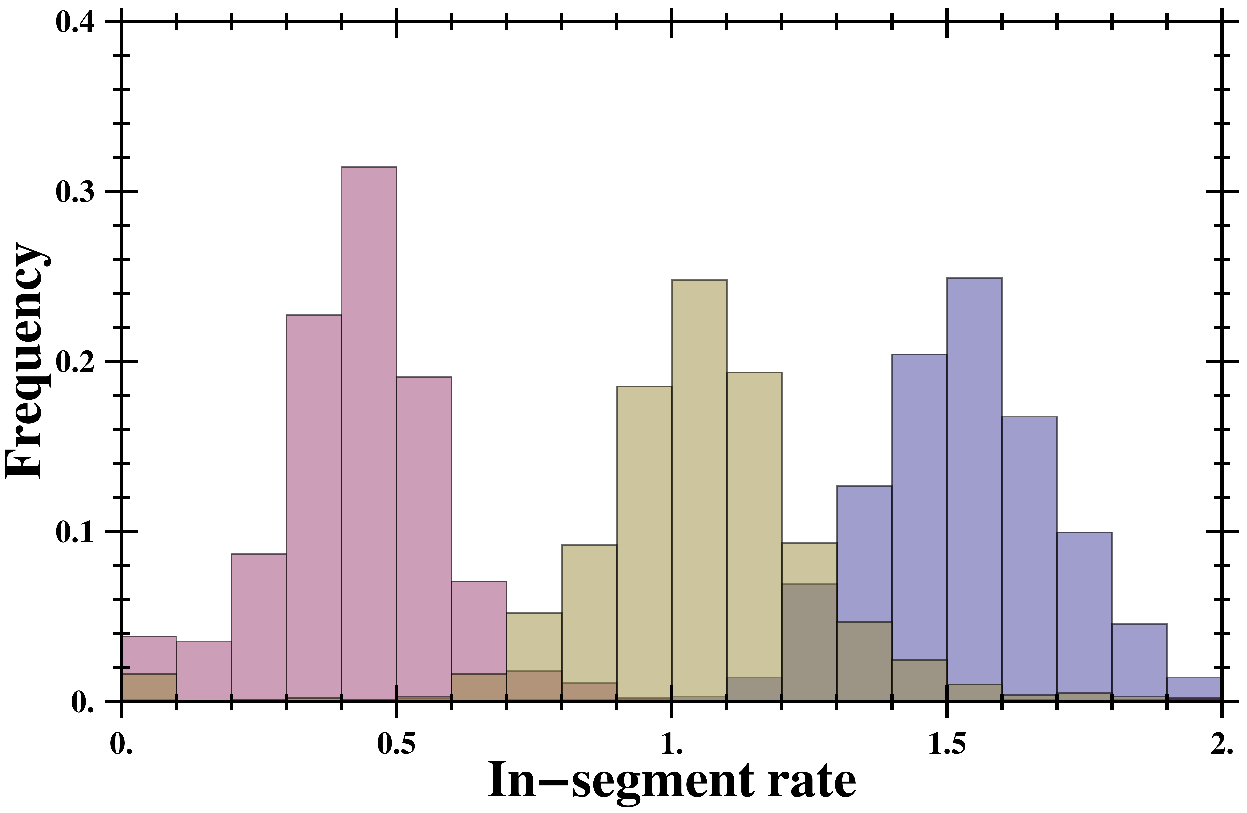}
	\end{subfigure}
	\caption{
		Top: Histogram of the boundaries locations. Bottom: Histogram of the in-segment mean rates. Both figures were calculated for a set of 2000 simulations similar to that shown in Fig. \ref{fig:simu1}.
	}
	\label{fig:analisis1}
\end{figure}

The regression in the example above fits very well with the rate curve
used in simulation.  But sometimes regressions result qualitatively
different to the rate curve, showing more or less breaking points,
even for data simulated in the same conditions.  This effect is due to
chance.  To show this issue, 2000 simulations with the same conditions
were performed.  In 992 of them, two breaking points were found.  In
the others, there were found zero (44), one (208), three (419), four
(155), five (73), six (41), and seven or more (68) breaking points.
For cases in which two breaking points were found, statistics of the
most likely boundaries locations and in-segment mean rates were
calculated.  Figure \ref{fig:analisis1} shows histograms of those
statistics.

Figure \ref{fig:analisis1} (top) shows histograms for boundaries
locations.  It is clear that the bigger the step, the smaller the
uncertainty on its location.  Figure \ref{fig:analisis1} (bottom) shows
histograms of in-segment rates.  It is clear that the greater the
rate, the smaller its relative statistical error.

\begin{figure}[]
	\captionsetup[subfigure]{aboveskip=0pt,belowskip=-8pt}
	\centering
	\begin{subfigure}[b]{0.8\columnwidth}
		\hspace{0.01cm} \includegraphics[height=0.7\columnwidth]{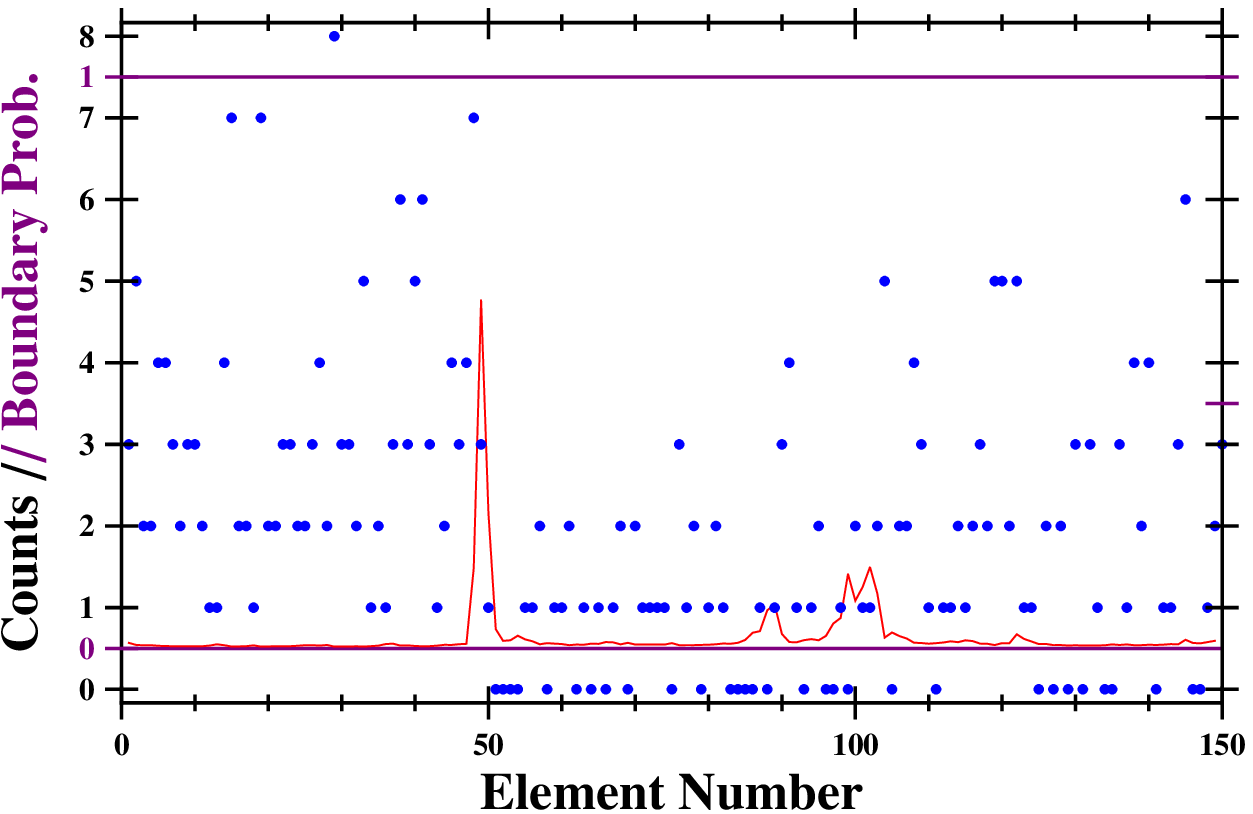}
	\end{subfigure}
        \vspace{0.2cm}
	
	\begin{subfigure}[b]{0.8\columnwidth}
		\includegraphics[height=0.7\columnwidth]{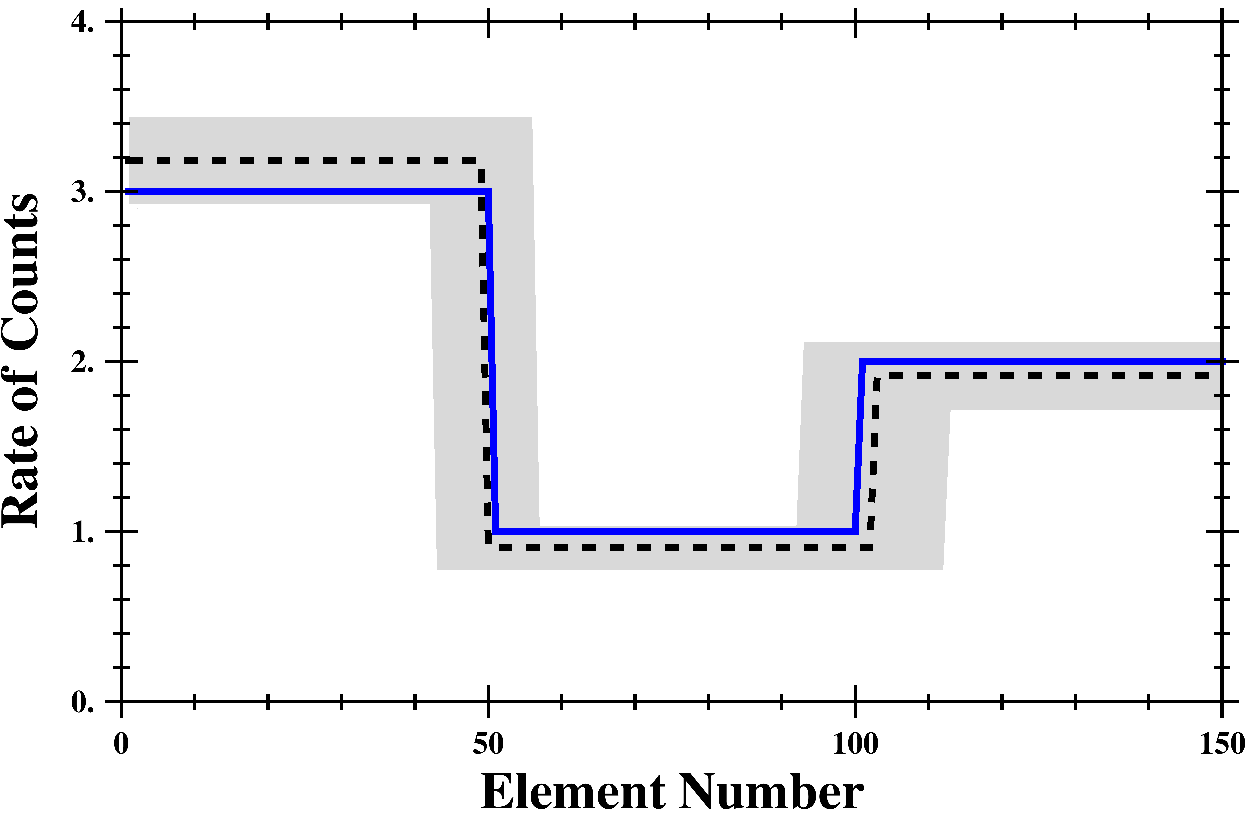}
	\end{subfigure}
	\caption{
		Top: Simulated data (blue dots) and boundaries location probability (red line).
		Bottom: Regression curve and its error estimation (black dashed line and gray zone), and the rate curve used in simulation (blue line).
	}
	\label{fig:simu2}
\end{figure}

Figure \ref{fig:simu2} shows data and boundaries location probabilities
for a simulation similar to the previous ones, but now with rates 3.0,
1.0 and 2.0.  Comparing Fig. \ref{fig:simu1} (top) and
Fig. \ref{fig:simu2} (top), we can see that in the latter one the
boundaries locations are found more accurately.

\begin{figure}[]
	\captionsetup[subfigure]{aboveskip=0pt,belowskip=-8pt}
	\centering
	\begin{subfigure}[b]{0.8\columnwidth}
		\hspace{0.01cm} \includegraphics[height=0.7\columnwidth]{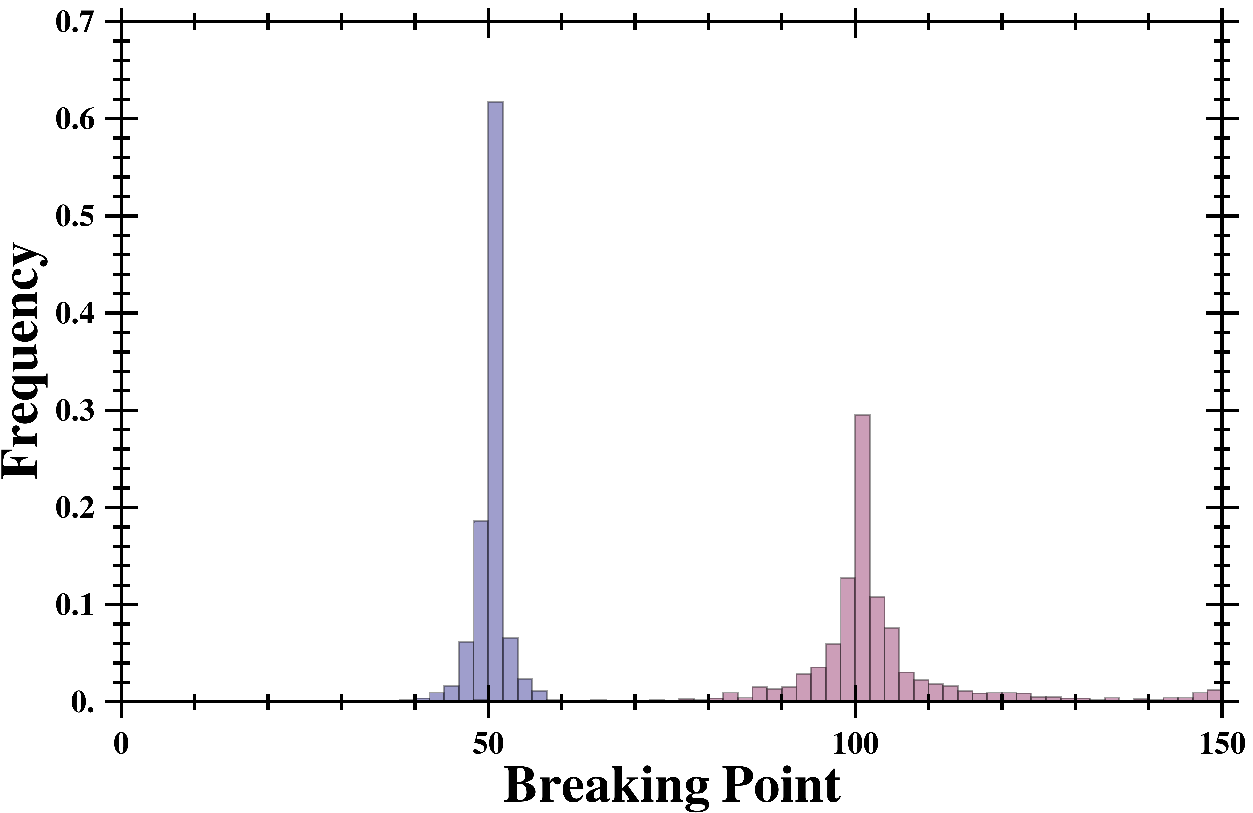}
	\end{subfigure}
       \vspace{0.2cm}
	
	\begin{subfigure}[b]{0.8\columnwidth}
		\includegraphics[height=0.7\columnwidth]{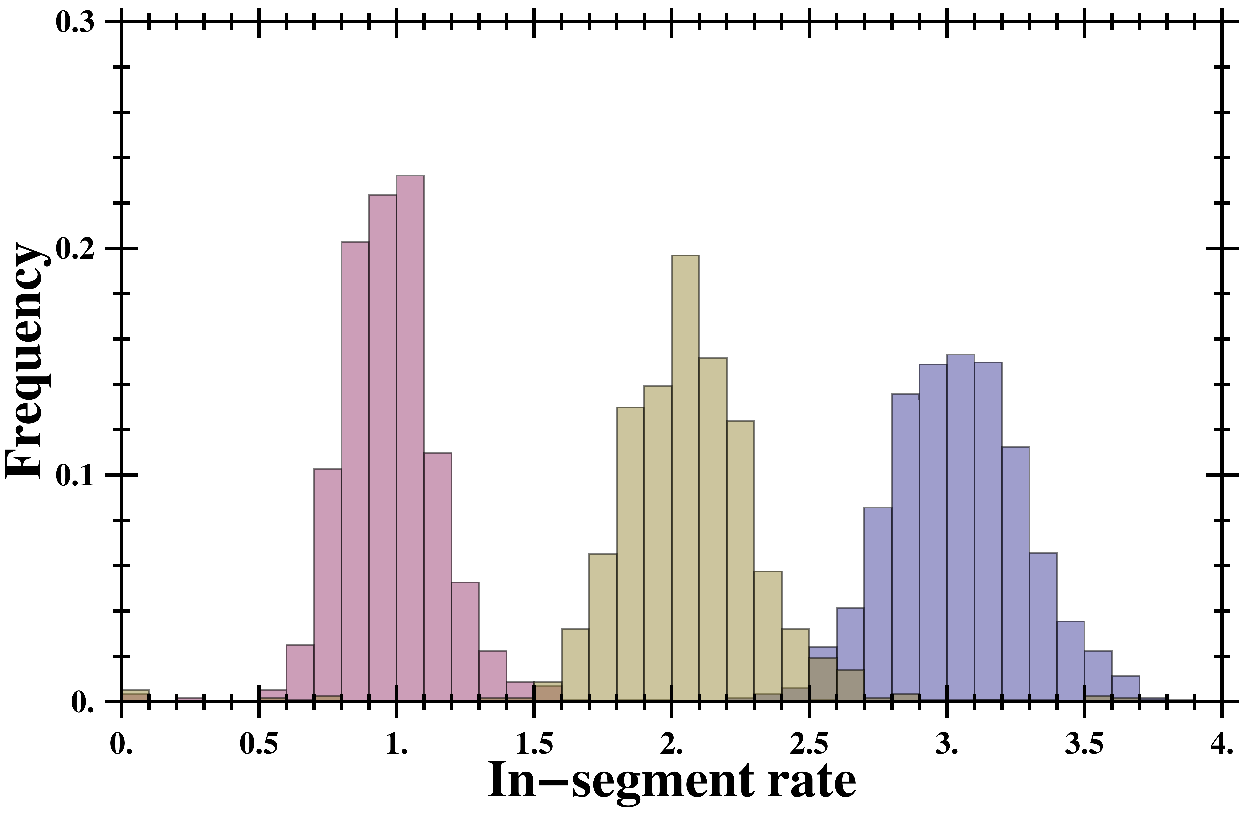}
	\end{subfigure}
	\caption{
		Top: Histogram of the boundaries locations. Bottom: Histogram of the in-segment rates. Both figures were calculated for a set of 2000 simulations similar to that shown in Fig. \ref{fig:simu2}.
	}
	\label{fig:analisis2}
\end{figure}

Again, 2000 simulations with the same conditions were performed.  In
1159 of them, two breaking points were found, while in the others, there
were found zero (1), one (33), three (499), four (167), five (75), six
(32), and seven or more (34) breaking points.  Figure
\ref{fig:analisis2} shows histograms of the statistics of the most
likely boundaries locations and in-segment rates for the simulations
with two breaking points found.  Comparing with Fig.
\ref{fig:analisis1}, we can see that the histograms are now narrower.
These results confirm what was stated above.

It is important to note that the probability for the \emph{real
  curve} 
to be \emph{completely} inside the region defined by the error
estimations of the regression is significantly less than one.  It is
easy to see why: if the errors were independent and equal to the standard error, the probability of satisfying $n$ error
conditions simultaneously would be
$0.68^n$.
But even the actual probability could be lower, since it is clear that
the errors must be dependent.  Nevertheless, the error estimations
presented here are useful to get an idea of the accuracy of the
regression.

Finally, the capability to detect a breaking point with this code 
was tested for different count rates.
To do this, simulated data sets of a single step in the count rate were used.
Data sets consist in 100 Poisson distributed elements,
the first 50 for a rate $r_1$ and the last 50 for a rate $r_2$.
1000 simulations were done for each pair $(r_1,r_2)$. Statistics of
successful detections are presented in Table \ref{tab:table1}.  A
successful detection is considered when only one breaking point between
elements 40 and 60 is detected.

Table \ref{tab:table1} shows that the smaller is the mean rate
difference and the smaller are the mean rates, the more difficult is
the detection of the step.  This result is expected because in a
Poisson distribution the variance is equal to the mean.

\begin{table}[]
  \begin{center}
    \caption{Statistics of successful detections for a single step.}
    \label{tab:table1}
    \begin{tabular}{c|ccccc}
      \toprule
	$r_1$\textbackslash $\,r_2$    & 3.0 & 2.0 & 1.6 & 1.2 & 0.8\\
      \midrule
	0.4 & 0.86 & 0.74 & 0.70 & 0.63 & 0.46\\
	0.8 & 0.77 & 0.70 & 0.60 & 0.36 & \\
	1.2 & 0.69 & 0.61 & 0.30 &     & \\
	1.6 & 0.65 & 0.28 &     &     & \\
	2.0 & 0.60 &     &     &     & \\
      \bottomrule
    \end{tabular}
  \end{center}
\end{table}

\section{Conclusions}

In this work, a code for Bayesian regression of piecewise constant
functions was adapted to handle data from Poisson processes.
For this purpose, equations for calculating the moments of the posteriors of segments of data
were found through Bayes theorem, considering the conjugate prior of the Poisson distribution as prior.
These results, as well as part of Hutter's method,
were used to calculate the most probable number of segments and their
boundaries.  Procedures for calculating in-segments mean rates and the
uncertainties of mean rates and boundaries locations are also
provided.  The resulting method is summarized in a code in
\emph{Mathematica}.

The code was applied to simulated data.
Firstly, two examples with tree segments were analyzed.
The code performed well in both cases considering the dispersion of data,
and the results improved in the case of higher mean rates and mean rates
differences.  This occurs because of the statistical dispersion of Poisson
distributed data, which is greater than the mean rate if the mean rate
is lower than one.

Finally, simulations of data of a single step were analyzed for
different rates, and statistics of the regressions with only one
breaking point are presented in a table.  This table shows the effect
of the rates and rate differences in the regression accuracy, and,
together with the errors estimations provided by the code, can serve
as an indicator of the reliability of the method.

Supplementary material including the source code for the algorithms can be found at the journal website \cite{supplementary}.

\begin{acknowledgements}
This work was partially supported by the National University of Rosario.
\end{acknowledgements}

\end{document}